\documentclass[sn-mathphys,iicol]{sn-jnl}

\usepackage{graphicx}
\usepackage{natbib}
\usepackage{doi}
\usepackage{siunitx}
\DeclareSIUnit{\electrons}{e\textsuperscript{$-$}}

\title{Picometer-precision few-tilt ptychotomography of 2D materials}

\author[1]{\fnm{Christoph} \sur{Hofer}}

\author[2]{\fnm{Kimmo} \sur{Mustonen}}

\author[2,3,4]{\fnm{Viera} \sur{Sk\'akalov\'a}}

\author[1]{\fnm{Timothy J.} \sur{Pennycook}}\email{timothy.pennycook@uantwerpen.be}

\affil[1]{\orgname{EMAT}, \orgdiv{University of Antwerp}, \orgaddress{\street{Campus Groenenborger}, \city{Antwerp}, \postcode{2020}, \country{Belgium}}}

\affil[2]{\orgdiv{University of Vienna}, \orgname{Faculty of Physics}, \orgaddress{\street{Boltzmanngasse 5}, \city{Vienna}, \postcode{1090}, \country{Austria}}}

\affil[3]{\orgdiv{Slovak Academy of Sciences}, \orgname{Institute of Electrical Engineering}, \orgaddress{\street{D\'ubravsk\'a cesta 3484/9}, \city{Bratislava}, \postcode{84104}, \country{Slovakia}}}

\affil[4]{\orgdiv{Danubia NanoTech, s.r.o.},  \orgaddress{\street{D\'ubravsk\'a cesta 3484/9}, \city{Bratislava}, \postcode{84104}, \country{Slovakia}}}

\hypersetup{
pdftitle={3D ptycho},
pdfsubject={},
pdfauthor={},
pdfkeywords={STEM, 2D materials, 3D structure, ptychography},
}

\begin{document}

\abstract{
From ripples to defects, edges and grain boundaries, the 3D atomic structure of 2D materials is critical to their properties. However the damage inflicted by conventional 3D analysis precludes its use with fragile 2D materials, particularly for the analysis of local defects. Here we dramatically increase the potential for precise local 3D atomic structure analysis of 2D materials, with both greatly improved dose efficiency and sensitivity to light elements.
We demonstrate light atoms can now be located in complex 2D materials with picometer precision at doses 30 times lower than previously possible. Moreover we demonstrate this using a material, WS$_2$, in which the light S atoms are in fact practically invisible to conventional methods.
 The key advance is combining the concept of few tilt tomography with highly dose efficient ptychography in scanning transmission electron microscopy. We further demonstrate the method experimentally with the even more challenging and newly discovered 2D CuI, leveraging a new extremely high temporal resolution camera.

}

\maketitle

Since the first isolation of graphene~\cite{Novoselov2004}, the outstanding properties of 2D materials have attracted huge attention. These properties emerge from their single atom or single unit cell thicknesses and make them promising candidates for a wide variety of applications including electronics, optoelectronics, energy storage devices, and ultra-sensitive detectors~\cite{Xiaoyan2016,Qijie2021}. However, despite their name, the properties of 2D materials are in fact determined by their 3D structures. For example, perfectly flat graphene is unstable~\cite{Mermin1968}. The famous stability of the material originates from its 3D structure~\cite{Meyer2007}. Such examples of the importance of the 3D structure of 2D materials are ubiquitous~\cite{Houmad2015,Ramasse2013,Zhang2016,Zhou2012,Hofer2019}. Not only do these materials easily deform out of plane to accommodate strain or defects, but they also frequently consist of three or more atomic layers, with this number increased multiple times in vertical heterostructures. It is thus no surprise that the 3D structure of 2D materials can dramatically alter their properties.

3D structures are typically determined with tomography. For atomic resolution this requires either atom probe tomography (APT) or electron tomography, but neither are well suited to 2D materials. APT requires samples be formed into sharp needles rather than 2D sheets~\cite{Kelly2007}, and conventional electron tomography ~\cite{Xu2015, Midgley2003} generally requires prohibitively large doses. The lengthy tilt series used by conventional tomography require a hundred times or more dose than a single image~\cite{Kak1988,Eder2014,Kotakoski2014}.
However, few tilt tomography allows one to determine the structure of 2D materials at far lower doses. Far fewer tilts are needed because it directly tracks each individual atom in the material as it is tilted. This has allowed the 3D structure of relatively robust defects in graphene to be determined with as few as two tilt angles~\cite{Hofer2018}, but most new 2D materials are more complicated than graphene, being both several atomic layers thick and containing a mix of elements widely spaced in atomic number. With such materials not all atoms are visible from every angle as they are in graphene. This is particularly true when relying on annular dark field (ADF) imaging, the mainstay for imaging 2D materials for the past decade. The sensitivity of the ADF signal to atomic number is extremely useful, allowing one to determine the spatial distribution of elements in materials, if sufficient signal can be obtained. However, the signal from light atoms is often completely obscured by the strong scattering of neighbouring heavy elements. This is especially true under low dose condtions due to the inefficiency of the ADF signal. Thus although the 3D structure of the transition metal dichalcogenide (TMD), MoS\textsubscript{2} has been possible with ADF tomography~\cite{Tian2020}, a more dose efficient method that is better able to locate more diverse combinations of light and heavy elements remains highly desirable. 

\begin{figure*}
	\center\includegraphics[width=0.95\textwidth]{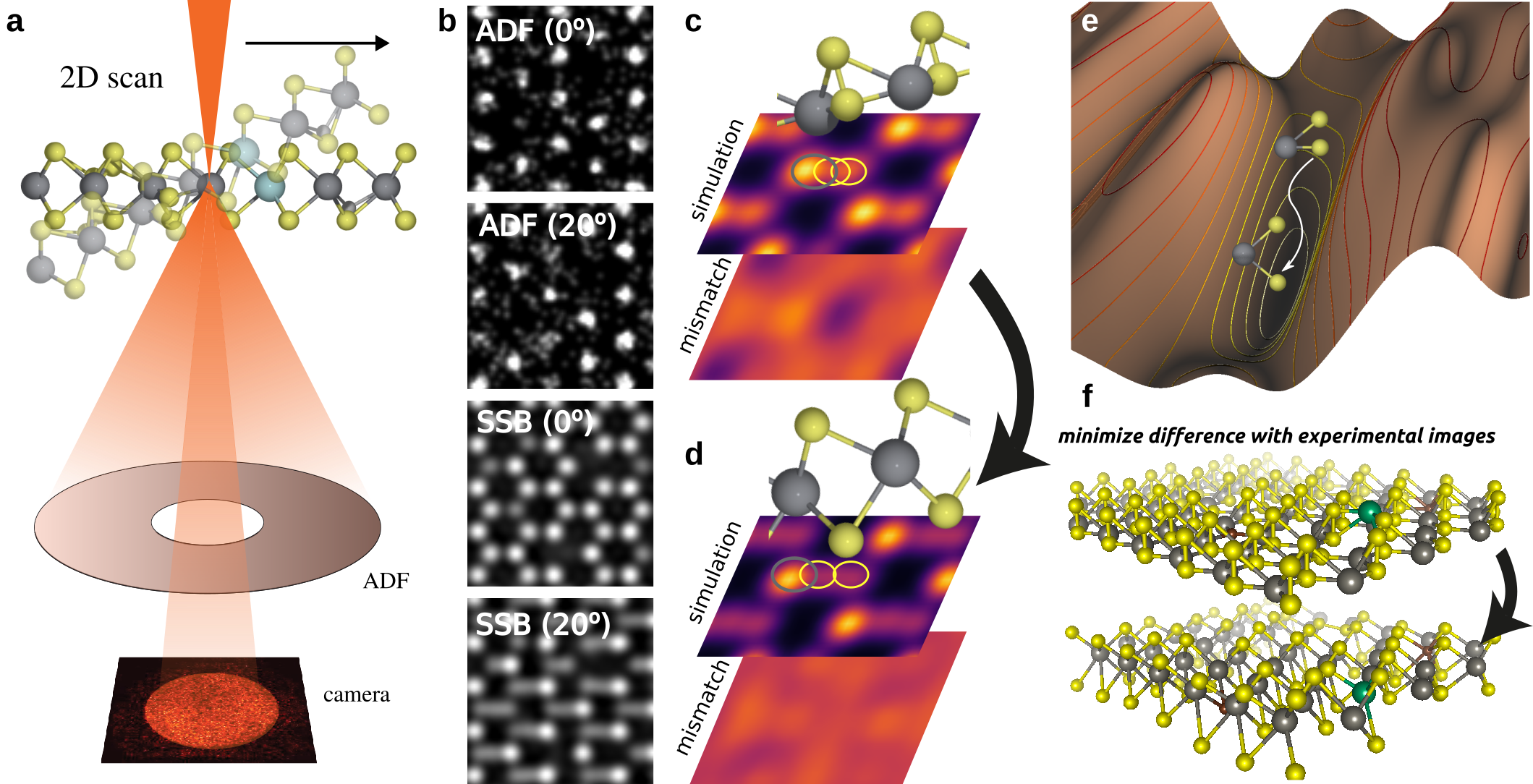}
	\caption{\label{fig:2Da} \textbf{Few tilt ptychotomography.} \textbf{a}, Schematic \textbf{b}, Simulated ADF and SSB images of WS\textsubscript{2} at a sample tilt of 0 and 20 degrees, respectively. \textbf{c,d,} Simulation of tilted models (wrong and correct S coordinates on c and d, respectively). The circles in the simulation indicate the projected positions and the mismatch is the difference between the simulation and the corresponding target image (experimental image). \textbf{e,} Correlation landscape of a single atom computed as the negative correlation between simulation and experiment. \textbf{f,} Initialized (top) and optimized (bottom) model.
	}	
\end{figure*}

Here we introduce a new form of electron tomography which provides all the benefits of ADF tomography but is able to simultaneously locate both heavy and light elements far more precisely and at much lower doses than previously possible. By combining ptychography with ADF imaging as in Fig.~\ref{fig:2Da}a, we achieve picometer precision solutions of the 3D structure of complex 2D materials at a small fraction of the dose previously required. With a thirty fold reduction in dose, the new method is able to solve the structure of WS$_2$ at the same precision as the best previously demonstrated for MoS$_2$, despite the fact that the S atoms are not even visible in WS$_2$ in ADF images (Fig.~\ref{fig:2Da}b). Although the use of ptychography provides the greatest step change in dose efficiency, it is still crucial to combine it with the Z-contrast signal for optimal use of the available information. This necessitates focused rather than defocused probe ptychography, which cannot provide a simultaneously acquired ADF signal. We demonstrate the method experimentally with a new complex 2D material, CuI, using an extremely high temporal resolution camera~\cite{jannis2021event} that enables us to perform very rapid low dose 4D STEM. This experimental ability provides the key to unlocking the connection between structure and function in new and emerging 2D materials, most of which are predicted to posses complex 3D structures \cite{mounet2018two}.

\subsection*{Few-tilt ptychotomography of WS\textsubscript{2}}
The experimental setup is illustrated in Figure~\ref{fig:2Da}a, and consists of obtaining simultaneous 4D STEM and ADF data at each tilt angle.  
We introduce the method with defective monolayer WS\textsubscript{2} data simulated at an electron dose of \SI{5000}{\electrons\per\angstrom\squared} per tilt angle. At this electron dose the heavy W atoms are clearly visible in the ADF images but the lighter S atoms are not (Fig.~\ref{fig:2Da}b). All the elements are however clearly visible in the ptychographic single side band (SSB) images. The vacancies and substitutional C and Nb atoms can also be distinguished (see Extended Data Fig.~\ref{fig:2D}). 





The 3D structure is solved by optimising the match between the input images and images simulated based on a model. Both, the model and the imaging conditions used in the simulations based on it are iteratively updated to maximise the correlation. 
An initial model with the correct lateral positions and elemental identification can typically be obtained from the untilted ADF and ptychographic images. The quantitative Z-contrast of the ADF signal is particularly beneficial to the elemental identification, even with very noisy images. Initial vertical positions can be assigned as flat or use prior knowledge. For the WS\textsubscript{2} we assigned an arbitrary offset of $z=\SI{1.5}{\angstrom}$ between the S atoms with one of the S planes aligned vertically with the W atoms as illustrated in Extended Data Fig.~\ref{fig:conv}. 
With the initial model established, optimisation is performed using the data from each tilt angle, iteratively updating the position of each atom. In addition to the tilt angle, the source-size broadening, linear drift, scan distortions and atomic scattering factors are included as adjustable parameters to obtain the best match.

As in all tomography, the shift of an atom in the projected image upon tilting is used to infer its 3D position. This shift is apparent in the ptychographic SSB input image shown in Figure ~\ref{fig:2Da}b with the sample tilted to 20$^\circ$, and in the SSB images simulated from the models shown in Fig.~\ref{fig:2Da}c. As the WS\textsubscript{2} model is optimised, the lower layer of S atoms extends further below the plane of the W atoms, causing the locations of the atoms to spread out laterally from the initial configuration in c to the final optimised positions in d. Maps of the mismatch between the ptychographic images from these configurations and the input image at this tilt are also displayed in parts c and d. With the lower S atom too close to the W plane, the projected W and S positions are close together and the mismatch is significant. As the lower S atom extends further below the W plane towards the position typical for TMDs, the projected positions of the atoms spreads out and the mismatch is reduced. 

The optimisation is further illustrated in Fig.~\ref{fig:2Da}e with a 3D rendering of the negative correlation of the match between the input and simulated data vs the position of the lower S atom in one of the WS\textsubscript{2} units. The correlation potential landscape of a single S atom was calculated by shifting it in all directions and tracking the merit function. A sharp valley appears where an optimisation algorithm can determine the correct atomic position by simple minimisation. Since the potential landscape depends on the other atoms of the system, the optimisation is iterated atom-by-atom in multiple cycles. When all atomic positions are converged, the final model shows an excellent match with the input model.

Using simulations and a predefined target model allows us to quantify the accuracy and precision of the method by calculating the difference between the target and optimised models. Our calculations show that the method can provide a precision in the low picometer range under low-dose conditions. We obtain a precision of \SI{4.8}{\pm} for the heavy W and Nb atoms at \SI{5e4}{\electrons\per\angstrom\squared}. As the heavier elements produce a stronger signal than lighter elements in the ADF images, they are able to be located with greater precision. The best precision previously achieved for S atoms in a TMD was \SI{15}{\pm} using a total dose of \SI{4.1e5}{\electrons\per\angstrom\squared} \cite{Tian2020} and 13 tilt angles. Our few tilt ptychotomography method achieves this specific precision with only 3 tilts and 1/30th the total dose despite WS$_2$ being a far more difficult material in which to locate light atoms than the MoS$_2$ used by Tian et al. Compared to the Mo in MoS$_2$, the near factor of two increase in atomic number of the W makes the S$_2$ atoms all but invisible in ADF images of WS$_2$, but the ptychography shows them clearly at this dose. The statistical relationship between the number of tilt angles and the precision is discussed in detail in the supplementary information for the full defected structure which includes very light C atoms.

\subsection*{Experimental example: 2D CuI}
\begin{figure}
	\center\includegraphics[width=0.49\textwidth]{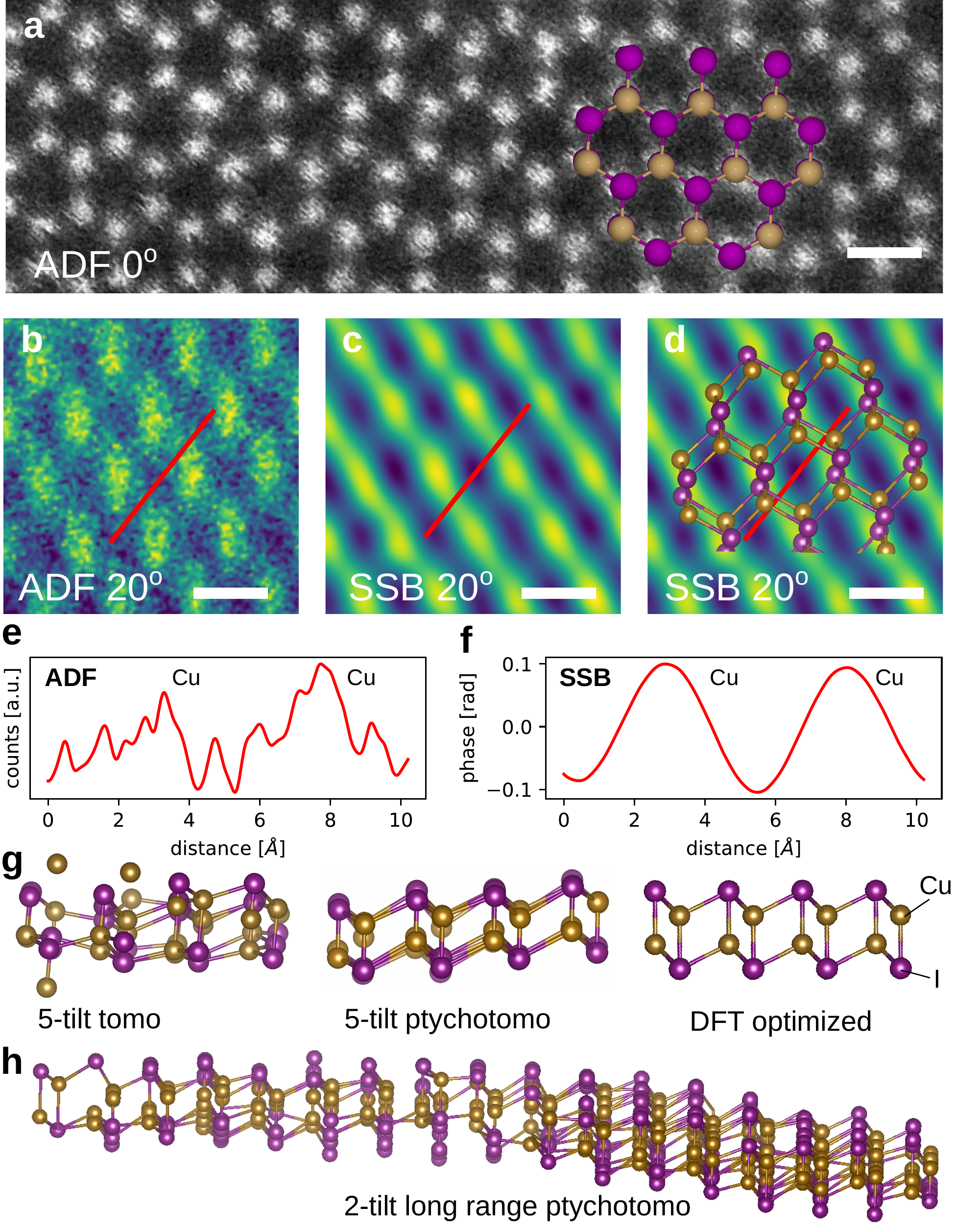}
	\caption{\label{fig:lp} \textbf{Characterization of 2D CuI.} \textbf{a,} High, \SI{1e6}{\electrons\per\angstrom\squared}, dose ADF image of 2D CuI taken at zero tilt angle.
	Simultaneous low dose \SI{5e4}{\electrons\per\angstrom\squared} \textbf{b,} ADF and \textbf{c,} SSB images taken at a 20$^\circ$ angle. 
\textbf{d,} Model obtained from ptychotomography is overlaid on the SSB image, showing how the tilt makes all the atoms visible. \textbf{e,f} Line profiles from the red lines in \textbf{b} and \textbf{c}, respectively. \textbf{g,} the 3D structures from 5 tilt tomography using just the ADF signal and from the combined ADF and SSB data are visualized side on alongside the results of DFT relaxation. \textbf{h} Two-tilt reconstruction of a larger area showing a significant undulation.
Scale bars are 0.5 nm.	}	
\end{figure}

We now apply the method to an experimental data set from monolayer hexagonal CuI. This new 2D material has very recently been discovered, and is stabilised between two graphene layers~\cite{mustonen2021exotic}. Structurally, it is equivalent to a single layer of $\beta$-CuI, with each atomic column containing one Cu and one I atom. Therefore the structure appears as a simple honeycomb lattice when untilted, as shown in the high \SI{1e6}{\electrons\per\angstrom\squared} dose  ADF image displayed in Fig.~\ref{fig:lp}a.

To reduce the dose imposed on the region of CuI used for the reconstruction we performed the few tilt series in a previously unirradiated region. We first acquired a series of 10 scans at zero tilt angle with a \SI{2}{\micro\second} dwell time and a dose of \SI{5e3}{\electrons\per\angstrom\squared} per scan. The simultaneous recording of ADF and 4D STEM data at this speed was made possible by our event driven Timepix3 detector~\cite{Poikela_2014,jannis2021event}. Importantly, the rapid scan speed greatly facilitates the low dose operation and greatly reduces the effects of drift in each scan. We build up signal by summing the drift corrected scans to obtain ADF and SSB images at \SI{5e4}{\electrons\per\angstrom\squared} with next to no motion blur. 

To obtain the full 3D configuration, we use data from the same region at specimen tilts of $\alpha=20^\circ$, $ \alpha=25^\circ$, $\alpha=30^\circ$, and $\beta=15^\circ$ again using drift corrected series of \SI{2} {\micro\second} dwell time \SI{5e4}{\electrons\per\angstrom\squared} scans. At this dose it is not possible to clearly determine the positions of the Cu atoms in the tilted projections from the ADF images alone. The high atomic number of iodine conceals the lighter Cu atoms in the ADF images as shown in Fig.\ \ref{fig:lp}b. The 3D structure thus cannot be accurately determined from the ADF images alone. Indeed, even at very high electron doses and utilising greater numbers of tilt angles the ADF images alone were not sufficient. The ptychography, however, provides a much stronger signal from which the projected positions of all the atoms can be more reliably extracted, as illustrated in the SSB image shown in Fig.\ \ref{fig:lp}c. 

The optimised model is overlaid on the same SSB image in Fig.\ \ref{fig:lp}d, illustrating how the tilt shifts the projected positions of the atoms laterally. Despite the signal from the Cu atoms not being fully separated from that of the I in the images, the well resolved elongated shape of the tilted columns is nonetheless sufficient for the tilt algorithm because it maximises the correlation between the simulated and experimental images, which does not require the atoms to be individually resolved. However the strength of the signal from the atoms is paramount. Line traces from along the red lines in Fig.\ \ref{fig:lp}b and c are shown in e and f. The strong smooth signal from the SSB ptychography is far superior for fitting atomic positions to than the noisy ADF image, and allows the algorithm to correctly identify the projected positions of even partially overlapping atoms. 

The enormous benefit provided by the ptychographic phase image for few tilt reconstruction is ultimately demonstrated by comparing the models produced by the algorithm using five tilts with and without it to a model relaxed by density functional theory (DFT). These are shown viewed from the same angle in Fig.\ \ref{fig:lp}g. Because of its relatively weak signal the ADF only reconstruction is far more disordered than that of the combined SSB and ADF reconstruction, which is a much closer match to the DFT model. Interestingly, the precision of the results is greatest when the optimisation algorithm includes both ADF and ptychographic signals, indicating that the additional information is still very useful even if it is noisy. Thus, there remains significant utility in obtaining the simultaneous ADF signal at each tilt angle, as is very easy to do in focused probe ptychography. 

Analysis of the z-heights output by the algorithm shows a standard deviation of \SI{12.8}{\pm} and \SI{14}{\pm} for iodine and copper respectively. Several factors likely contribute to this including residual non-linear drift, scan distortions, the finite accuracy of the tilting holder and even out-of-plane atomic vibrations. However these standard deviations do not take into account the possibility that the structure might actually be curved. 
Fig.\ \ref{fig:lp}h shows an example of a reconstruction from a wider area where the 2D CuI flake is indeed not perfectly flat, but rather slightly deformed. We were only able to collect data from two tilts before the area contaminated, but the precision is already sufficient to detect a low-frequency undulation in the CuI structure. This can be directly detected in the tilted projection, where the curvature leads to a slight distortion in the lattice (see Extended Data Fig.\ \ref{Sfig:curve}). There are various possible reasons for the non-flatness of 2D CuI such as the deformation of the encapsulating graphene layers, strain, contamination, and defects in the structure. In this example, we attribute the undulation to a small pore nearby.

\subsection*{Discussion and Outlook}
We have developed and experimentally demonstrated a new few-tilt tomographic approach that combines ADF and ptychographic imaging to greatly enhance our ability to solve the 3D structure of 2D materials. The method works as long as the position of each atom can be tracked in all tilts, even if they are not fully resolved. This prerequisite is fulfilled in all foreseeable 2D materials, including those which are structurally complex, as is predicted to be the case for most future 2D materials \cite{mounet2018two}. The addition of SSB ptychography provides far greater dose efficiency imaging that allows a more accurate and precise determination of atomic positions at significantly lower doses than with ADF imaging alone. Importantly, the ptychography also provides far greater sensitivity to light atoms near heavy elements that are hidden in the ADF signal. With the greater ability to efficiently detect all atoms, we require fewer tilt angles to accurately determine the height of each atom. From simulations, it is clear that accurate results can be obtained from very few tilt angles indeed with this method, further facilitating the use of minimal dose. We showed that although as few as two tilt angles can provide an accurate reconstruction for multiple atomic layer 2D materials such as WS$_2$, increasing the number of tilt angles  increases the precision. The maximum number of tilt angles depends on the stability of the material. The precision is also influenced by the cross-sections of the elements in the system as shown for W, S, Cu and I. Experimentally we demonstrated the technique with the newly discovered 2D CuI material, which has an even more complicated four atomic layer thick structure. With a total of five tilts we achieve a 3D reconstruction that is in excellent agreement with the results of DFT with our new method, an achievement which is not possible with ADF imaging alone, although the greatest precision is obtained when the ADF data is included in the optimization. Crucial to the experimental precision was our use of an event driven camera which enabled us to perform multiple scans at $\mu$s dwell time which were then drift corrected and summed at each tilt to minimise drift. The results of the work open up a new route for precise and accurate determination of the complete 3D atomic network of thin beam sensitive materials. The method will be especially useful when identifying structures such as defects in 2D materials, which are usually only stable for at most a few scans until structural changes occur.

\section*{Methods}
\subsection*{Optimization}
In detail, the correlation function to be maximized is
\begin{equation}
\resizebox{.9\hsize}{!}{$ R_{tot} =  \sum\limits_V \sum\limits_{m}\sum\limits_{i=1}^N\frac{\left( \mu^{sim,m,V}-I^{sim,m,V}_i\right) \left(\mu^{exp,m,V}-I^{exp,m,V}_i\right) }{\sigma^{sim,m,V}\sigma^{exp,m,V}\left(N-1\right)}, $}
\end{equation}
where the sum runs over all $N$ probe positions in the scan, and $I^{exp,m,V}_{i}$ and $I^{sim,m,V}_{i}$ are the intensities of the $i$th pixel of the experimental data and the numerical simulation of the imaging method $m$ (ADF or SSB) from a view V, respectively. 
$\mu^{exp,m,V}$ and $\mu^{sim,m,V}$ are the corresponding mean values of the experimental and the simulated images, respectively and $\sigma^{exp,m,V}$ and $\sigma^{sim,m,V}$ are the corresponding standard deviations of the experimental and simulated images, respectively. This function is maximised based on quadratic interpolation: Here, the gradients in each direction are calculated by a finite difference method and the minimum of a quadratic fit is estimated and used as the next iteration value. 
The whole $R_{tot}$ can be interpreted as the sum of the correlation of each image input to the algorithm.

\subsection*{Electron microscopy}

$\mu$s dwell time electron ptychography was conducted using a fast, event-driven Timepix3~\cite{Poikela_2014,jannis2021event} camera in a probe-corrected FEI Themis Z instrument with a probe convergence angle of 30~mrad and a beam current of 1\ pA. 10 scans of $1024\times1024$ probe positions were acquired sequentially at each tilt angle ($0^\circ, \alpha=20^\circ,\alpha=30^\circ$ and $\beta=15^\circ$) with a dwell time of \SI{2}{\micro\second}. All 10 data sets were processed with the SSB method and the final images were aligned and averaged using non-rigid registration. The algorithms for the SSB and the 3D reconstruction written in python can be found on gitlab~\cite{gitlab}. 

\subsection*{STEM simulations}
STEM image multislice simulations were carried out using the PyQSTEM Package~\cite{SUSI2019Efficient} using a 60~kV accelerating voltage, a convergence angle of 30~mrad and a probe step size of 18~pm. 3 slices were used for the normal incident image, and up to 30 slices for the tilted structures. Thermal vibrations and the finite source size are taken into account with a Gaussian broadening of \SI{1}{\angstrom}

During the optimisation procedure, simulations using the convolution method were used for its computationally efficiency~\cite{Kirkland2010}. The accuracy of this method is discussed in the supplementary information.

\section*{Acknowledgement}
C.H. and T.J.P. acknowledge funding from the European Research Council (ERC) via grant number 802123-HDEM.
V.S. was supported by the Austrian Science Fund (FWF) (project no. I2344-N36),  the Slovak Research and Development Agency (APVV-16-0319), the project CEMEA of the Slovak Academy of Sciences, ITMS project code 313021T081 of the Research \& Innovation Operational Programme and from the V4-Japan Joint Research Program (BGapEng).
Danubia NanoTech s.r.o. has received funding from the European Union's Horizon 2020 research and innovation programme under grant agreement No 101008099 (CompSafeNano) and also thanks Mr. Kamil Bern\'ath for his support. 

\bibliography{references}  


\begin{thebibliography}{28}
\ifx \bisbn   \undefined \def \bisbn  #1{ISBN #1}\fi
\ifx \binits  \undefined \def \binits#1{#1}\fi
\ifx \bauthor  \undefined \def \bauthor#1{#1}\fi
\ifx \batitle  \undefined \def \batitle#1{#1}\fi
\ifx \bjtitle  \undefined \def \bjtitle#1{#1}\fi
\ifx \bvolume  \undefined \def \bvolume#1{\textbf{#1}}\fi
\ifx \byear  \undefined \def \byear#1{#1}\fi
\ifx \bissue  \undefined \def \bissue#1{#1}\fi
\ifx \bfpage  \undefined \def \bfpage#1{#1}\fi
\ifx \blpage  \undefined \def \blpage #1{#1}\fi
\ifx \burl  \undefined \def \burl#1{\textsf{#1}}\fi
\ifx \doiurl  \undefined \def \doiurl#1{\url{https://doi.org/#1}}\fi
\ifx \betal  \undefined \def \betal{\textit{et al.}}\fi
\ifx \binstitute  \undefined \def \binstitute#1{#1}\fi
\ifx \binstitutionaled  \undefined \def \binstitutionaled#1{#1}\fi
\ifx \bctitle  \undefined \def \bctitle#1{#1}\fi
\ifx \beditor  \undefined \def \beditor#1{#1}\fi
\ifx \bpublisher  \undefined \def \bpublisher#1{#1}\fi
\ifx \bbtitle  \undefined \def \bbtitle#1{#1}\fi
\ifx \bedition  \undefined \def \bedition#1{#1}\fi
\ifx \bseriesno  \undefined \def \bseriesno#1{#1}\fi
\ifx \blocation  \undefined \def \blocation#1{#1}\fi
\ifx \bsertitle  \undefined \def \bsertitle#1{#1}\fi
\ifx \bsnm \undefined \def \bsnm#1{#1}\fi
\ifx \bsuffix \undefined \def \bsuffix#1{#1}\fi
\ifx \bparticle \undefined \def \bparticle#1{#1}\fi
\ifx \barticle \undefined \def \barticle#1{#1}\fi
\bibcommenthead
\ifx \bconfdate \undefined \def \bconfdate #1{#1}\fi
\ifx \botherref \undefined \def \botherref #1{#1}\fi
\ifx \url \undefined \def \url#1{\textsf{#1}}\fi
\ifx \bchapter \undefined \def \bchapter#1{#1}\fi
\ifx \bbook \undefined \def \bbook#1{#1}\fi
\ifx \bcomment \undefined \def \bcomment#1{#1}\fi
\ifx \oauthor \undefined \def \oauthor#1{#1}\fi
\ifx \citeauthoryear \undefined \def \citeauthoryear#1{#1}\fi
\ifx \endbibitem  \undefined \def \endbibitem {}\fi
\ifx \bconflocation  \undefined \def \bconflocation#1{#1}\fi
\ifx \arxivurl  \undefined \def \arxivurl#1{\textsf{#1}}\fi
\csname PreBibitemsHook\endcsname

\bibitem{Novoselov2004}
\begin{barticle}
\bauthor{\bsnm{Novoselov}, \binits{K.S.}},
\bauthor{\bsnm{Geim}, \binits{A.K.}},
\bauthor{\bsnm{Morozov}, \binits{S.V.}},
\bauthor{\bsnm{Jiang}, \binits{D.}},
\bauthor{\bsnm{Zhang}, \binits{Y.}},
\bauthor{\bsnm{Dubonos}, \binits{S.V.}},
\bauthor{\bsnm{Grigorieva}, \binits{I.V.}},
\bauthor{\bsnm{Firsov}, \binits{A.A.}}:
\batitle{{Electric Field Effect in Atomically Thin Carbon Films}}.
\bjtitle{Science}
\bvolume{306}(\bissue{5696}),
\bfpage{666}--\blpage{669}
(\byear{2004}).
\doiurl{10.1126/SCIENCE.1102896}
\end{barticle}
\endbibitem

\bibitem{Xiaoyan2016}
\begin{barticle}
\bauthor{\bsnm{Zhang}, \binits{X.}},
\bauthor{\bsnm{Hou}, \binits{L.}},
\bauthor{\bsnm{Ciesielski}, \binits{A.}},
\bauthor{\bsnm{Samorì}, \binits{P.}}:
\batitle{2d materials beyond graphene for high-performance energy storage
  applications}.
\bjtitle{Advanced Energy Materials}
\bvolume{6}(\bissue{23}),
\bfpage{1600671}
(\byear{2016})
{\href{https://arxiv.org/abs/https://onlinelibrary.wiley.com/doi/pdf/10.1002/aenm.201600671}{{https://onlinelibrary.wiley.com/doi/pdf/10.1002/aenm.201600671}}}.
\doiurl{10.1002/aenm.201600671}
\end{barticle}
\endbibitem

\bibitem{Qijie2021}
\begin{barticle}
\bauthor{\bsnm{Ma}, \binits{Q.}},
\bauthor{\bsnm{Ren}, \binits{G.}},
\bauthor{\bsnm{Xu}, \binits{K.}},
\bauthor{\bsnm{Ou}, \binits{J.Z.}}:
\batitle{Tunable optical properties of 2d materials and their applications}.
\bjtitle{Advanced Optical Materials}
\bvolume{9}(\bissue{2}),
\bfpage{2001313}
(\byear{2021})
{\href{https://arxiv.org/abs/https://onlinelibrary.wiley.com/doi/pdf/10.1002/adom.202001313}{{https://onlinelibrary.wiley.com/doi/pdf/10.1002/adom.202001313}}}.
\doiurl{10.1002/adom.202001313}
\end{barticle}
\endbibitem

\bibitem{Mermin1968}
\begin{barticle}
\bauthor{\bsnm{Mermin}, \binits{N.D.}}:
\batitle{Crystalline order in two dimensions}.
\bjtitle{Phys. Rev.}
\bvolume{176},
\bfpage{250}--\blpage{254}
(\byear{1968}).
\doiurl{10.1103/PhysRev.176.250}
\end{barticle}
\endbibitem

\bibitem{Meyer2007}
\begin{barticle}
\bauthor{\bsnm{Meyer}, \binits{J.C.}},
\bauthor{\bsnm{Geim}, \binits{A.K.}},
\bauthor{\bsnm{Katsnelson}, \binits{M.I.}},
\bauthor{\bsnm{Novoselov}, \binits{K.S.}},
\bauthor{\bsnm{Booth}, \binits{T.J.}},
\bauthor{\bsnm{Roth}, \binits{S.}}:
\batitle{{The structure of suspended graphene sheets}}.
\bjtitle{Nature}
\bvolume{446}(\bissue{7131}),
\bfpage{60}--\blpage{63}
(\byear{2007}).
\doiurl{10.1038/nature05545}
\end{barticle}
\endbibitem

\bibitem{Houmad2015}
\begin{barticle}
\bauthor{\bsnm{Houmad}, \binits{M.}},
\bauthor{\bsnm{Zaari}, \binits{H.}},
\bauthor{\bsnm{Benyoussef}, \binits{A.}},
\bauthor{\bsnm{{El Kenz}}, \binits{A.}},
\bauthor{\bsnm{Ez-Zahraouy}, \binits{H.}}:
\batitle{{Optical conductivity enhancement and band gap opening with silicon
  doped graphene}}.
\bjtitle{Carbon}
\bvolume{94},
\bfpage{1021}--\blpage{1027}
(\byear{2015}).
\doiurl{10.1016/J.CARBON.2015.07.033}
\end{barticle}
\endbibitem

\bibitem{Ramasse2013}
\begin{barticle}
\bauthor{\bsnm{Ramasse}, \binits{Q.M.}},
\bauthor{\bsnm{Seabourne}, \binits{C.R.}},
\bauthor{\bsnm{Kepaptsoglou}, \binits{D.-M.}},
\bauthor{\bsnm{Zan}, \binits{R.}},
\bauthor{\bsnm{Bangert}, \binits{U.}},
\bauthor{\bsnm{Scott}, \binits{A.J.}}:
\batitle{{Probing the Bonding and Electronic Structure of Single Atom Dopants
  in Graphene with Electron Energy Loss Spectroscopy}}.
\bjtitle{Nano Letters}
\bvolume{13}(\bissue{10}),
\bfpage{4989}--\blpage{4995}
(\byear{2013}).
\doiurl{10.1021/nl304187e}
\end{barticle}
\endbibitem

\bibitem{Zhang2016}
\begin{barticle}
\bauthor{\bsnm{Zhang}, \binits{S.J.}},
\bauthor{\bsnm{Lin}, \binits{S.S.}},
\bauthor{\bsnm{Li}, \binits{X.Q.}},
\bauthor{\bsnm{Liu}, \binits{X.Y.}},
\bauthor{\bsnm{Wu}, \binits{H.A.}},
\bauthor{\bsnm{Xu}, \binits{W.L.}},
\bauthor{\bsnm{Wang}, \binits{P.}},
\bauthor{\bsnm{Wu}, \binits{Z.Q.}},
\bauthor{\bsnm{Zhong}, \binits{H.K.}},
\bauthor{\bsnm{Xu}, \binits{Z.J.}}:
\batitle{{Opening the band gap of graphene through silicon doping for the
  improved performance of graphene/GaAs heterojunction solar cells}}.
\bjtitle{Nanoscale}
\bvolume{8}(\bissue{1}),
\bfpage{226}--\blpage{232}
(\byear{2016}).
\doiurl{10.1039/C5NR06345K}
\end{barticle}
\endbibitem

\bibitem{Zhou2012}
\begin{barticle}
\bauthor{\bsnm{Zhou}, \binits{W.}},
\bauthor{\bsnm{Kapetanakis}, \binits{M.D.}},
\bauthor{\bsnm{Prange}, \binits{M.P.}},
\bauthor{\bsnm{Pantelides}, \binits{S.T.}},
\bauthor{\bsnm{Pennycook}, \binits{S.J.}},
\bauthor{\bsnm{Idrobo}, \binits{J.-C.}}:
\batitle{Direct determination of the chemical bonding of individual impurities
  in graphene}.
\bjtitle{Phys. Rev. Lett.}
\bvolume{109},
\bfpage{206803}
(\byear{2012}).
\doiurl{10.1103/PhysRevLett.109.206803}
\end{barticle}
\endbibitem

\bibitem{Hofer2019}
\begin{barticle}
\bauthor{\bsnm{Hofer}, \binits{C.}},
\bauthor{\bsnm{Skakalova}, \binits{V.}},
\bauthor{\bsnm{Monazam}, \binits{M.R.A.}},
\bauthor{\bsnm{Mangler}, \binits{C.}},
\bauthor{\bsnm{Kotakoski}, \binits{J.}},
\bauthor{\bsnm{Susi}, \binits{T.}},
\bauthor{\bsnm{Meyer}, \binits{J.C.}}:
\batitle{{Direct visualization of the 3D structure of silicon impurities in
  graphene}}.
\bjtitle{Applied Physics Letters}
\bvolume{114}(\bissue{5}),
\bfpage{53102}
(\byear{2019}).
\doiurl{10.1063/1.5063449}
\end{barticle}
\endbibitem

\bibitem{Kelly2007}
\begin{barticle}
\bauthor{\bsnm{Kelly}, \binits{T.F.}},
\bauthor{\bsnm{Miller}, \binits{M.K.}}:
\batitle{Atom probe tomography}.
\bjtitle{Review of Scientific Instruments}
\bvolume{78}(\bissue{3}),
\bfpage{031101}
(\byear{2007})
{\href{https://arxiv.org/abs/https://doi.org/10.1063/1.2709758}{{https://doi.org/10.1063/1.2709758}}}.
\doiurl{10.1063/1.2709758}
\end{barticle}
\endbibitem

\bibitem{Xu2015}
\begin{barticle}
\bauthor{\bsnm{Xu}, \binits{R.}},
\bauthor{\bsnm{Chen}, \binits{C.-C.}},
\bauthor{\bsnm{Wu}, \binits{L.}},
\bauthor{\bsnm{Scott}, \binits{M.C.}},
\bauthor{\bsnm{Theis}, \binits{W.}},
\bauthor{\bsnm{Ophus}, \binits{C.}},
\bauthor{\bsnm{Bartels}, \binits{M.}},
\bauthor{\bsnm{Yang}, \binits{Y.}},
\bauthor{\bsnm{Ramezani-Dakhel}, \binits{H.}},
\bauthor{\bsnm{Sawaya}, \binits{M.R.}},
\bauthor{\bsnm{Heinz}, \binits{H.}},
\bauthor{\bsnm{Marks}, \binits{L.D.}},
\bauthor{\bsnm{Ercius}, \binits{P.}},
\bauthor{\bsnm{Miao}, \binits{J.}}:
\batitle{{Three-dimensional coordinates of individual atoms in materials
  revealed by electron tomography}}.
\bjtitle{Nature Materials}
\bvolume{14}(\bissue{11}),
\bfpage{1099}--\blpage{1103}
(\byear{2015}).
\doiurl{10.1038/nmat4426}
\end{barticle}
\endbibitem

\bibitem{Midgley2003}
\begin{bchapter}
\bauthor{\bsnm{Midgley}, \binits{P.A.}},
\bauthor{\bsnm{Weyland}, \binits{M.}}:
\bctitle{{3D electron microscopy in the physical sciences: The development of
  Z-contrast and EFTEM tomography}}.
In: \bbtitle{Ultramicroscopy}
(\byear{2003}).
\doiurl{10.1016/S0304-3991(03)00105-0}
\end{bchapter}
\endbibitem

\bibitem{Kak1988}
\begin{bbook}
\bauthor{\bsnm{Kak}, \binits{A.C.}},
\bauthor{\bsnm{Slaney}, \binits{M.}},
\bauthor{\bsnm{{IEEE Engineering in Medicine}}},
\bauthor{\bsnm{Society.}, \binits{B.}}:
\bbtitle{{Principles of Computerized Tomographic Imaging}}.
\bpublisher{IEEE Press},
\blocation{New York}
(\byear{1988}).
\burl{https://books.google.at/books/about/Principles_of_Computerized_Tomographic_I.html?id=RKhrAAAAMAAJ&redir_esc=y}
\end{bbook}
\endbibitem

\bibitem{Eder2014}
\begin{barticle}
\bauthor{\bsnm{Eder}, \binits{F.R.}},
\bauthor{\bsnm{Kotakoski}, \binits{J.}},
\bauthor{\bsnm{Kaiser}, \binits{U.}},
\bauthor{\bsnm{Meyer}, \binits{J.C.}}:
\batitle{{A journey from order to disorder - atom by atom transformation from
  graphene to a 2D carbon glass.}}
\bjtitle{Scientific reports}
\bvolume{4},
\bfpage{4060}
(\byear{2014}).
\doiurl{10.1038/srep04060}
\end{barticle}
\endbibitem

\bibitem{Kotakoski2014}
\begin{barticle}
\bauthor{\bsnm{Kotakoski}, \binits{J.}},
\bauthor{\bsnm{Mangler}, \binits{C.}},
\bauthor{\bsnm{Meyer}, \binits{J.C.}}:
\batitle{{Imaging atomic-level random walk of a point defect in graphene}}.
\bjtitle{Nature Communications}
\bvolume{5}(\bissue{1}),
\bfpage{3991}
(\byear{2014}).
\doiurl{10.1038/ncomms4991}
\end{barticle}
\endbibitem

\bibitem{Hofer2018}
\begin{botherref}
\oauthor{\bsnm{Hofer}, \binits{C.}},
\oauthor{\bsnm{Kramberger}, \binits{C.}},
\oauthor{\bsnm{Monazam}, \binits{M.R.A.P.}},
\oauthor{\bsnm{Mangler}, \binits{C.}},
\oauthor{\bsnm{Mittelberger}, \binits{A.}},
\oauthor{\bsnm{Argentero}, \binits{G.}},
\oauthor{\bsnm{Kotakoski}, \binits{J.}},
\oauthor{\bsnm{Meyer}, \binits{J.C.}}:
Revealing the 3d structure of graphene defects.
2D Materials
\textbf{5}
(2018).
\doiurl{10.1088/2053-1583/aaded7}
\end{botherref}
\endbibitem

\bibitem{Tian2020}
\begin{barticle}
\bauthor{\bsnm{Tian}, \binits{X.}},
\bauthor{\bsnm{Kim}, \binits{D.S.}},
\bauthor{\bsnm{Yang}, \binits{S.}},
\bauthor{\bsnm{Ciccarino}, \binits{C.J.}},
\bauthor{\bsnm{Gong}, \binits{Y.}},
\bauthor{\bsnm{Yang}, \binits{Y.}},
\bauthor{\bsnm{Yang}, \binits{Y.}},
\bauthor{\bsnm{Duschatko}, \binits{B.}},
\bauthor{\bsnm{Yuan}, \binits{Y.}},
\bauthor{\bsnm{Ajayan}, \binits{P.M.}},
\bauthor{\bsnm{Idrobo}, \binits{J.C.}},
\bauthor{\bsnm{Narang}, \binits{P.}},
\bauthor{\bsnm{Miao}, \binits{J.}}:
\batitle{Correlating the three-dimensional atomic defects and electronic
  properties of two-dimensional transition metal dichalcogenides}.
\bjtitle{Nature Materials}
\bvolume{19},
\bfpage{867}--\blpage{873}
(\byear{2020}).
\doiurl{10.1038/s41563-020-0636-5}
\end{barticle}
\endbibitem

\bibitem{jannis2021event}
\begin{barticle}
\bauthor{\bsnm{Jannis}, \binits{D.}},
\bauthor{\bsnm{Hofer}, \binits{C.}},
\bauthor{\bsnm{Gao}, \binits{C.}},
\bauthor{\bsnm{Xie}, \binits{X.}},
\bauthor{\bsnm{Béché}, \binits{A.}},
\bauthor{\bsnm{Pennycook}, \binits{T.J.}},
\bauthor{\bsnm{Verbeeck}, \binits{J.}}:
\batitle{Event driven 4d stem acquisition with a timepix3 detector: Microsecond
  dwell time and faster scans for high precision and low dose applications}.
\bjtitle{Ultramicroscopy}
\bvolume{233},
\bfpage{113423}
(\byear{2022}).
\doiurl{10.1016/j.ultramic.2021.113423}
\end{barticle}
\endbibitem

\bibitem{mounet2018two}
\begin{barticle}
\bauthor{\bsnm{Mounet}, \binits{N.}},
\bauthor{\bsnm{Gibertini}, \binits{M.}},
\bauthor{\bsnm{Schwaller}, \binits{P.}},
\bauthor{\bsnm{Campi}, \binits{D.}},
\bauthor{\bsnm{Merkys}, \binits{A.}},
\bauthor{\bsnm{Marrazzo}, \binits{A.}},
\bauthor{\bsnm{Sohier}, \binits{T.}},
\bauthor{\bsnm{Castelli}, \binits{I.E.}},
\bauthor{\bsnm{Cepellotti}, \binits{A.}},
\bauthor{\bsnm{Pizzi}, \binits{G.}}, \betal:
\batitle{Two-dimensional materials from high-throughput computational
  exfoliation of experimentally known compounds}.
\bjtitle{Nature Nanotechnology}
\bvolume{13}(\bissue{3}),
\bfpage{246}--\blpage{252}
(\byear{2018})
\end{barticle}
\endbibitem

\bibitem{mustonen2021exotic}
\begin{barticle}
\bauthor{\bsnm{Mustonen}, \binits{K.}},
\bauthor{\bsnm{Hofer}, \binits{C.}},
\bauthor{\bsnm{Kotrusz}, \binits{P.}},
\bauthor{\bsnm{Markevich}, \binits{A.}},
\bauthor{\bsnm{Hulman}, \binits{M.}},
\bauthor{\bsnm{Mangler}, \binits{C.}},
\bauthor{\bsnm{Susi}, \binits{T.}},
\bauthor{\bsnm{Pennycook}, \binits{T.J.}},
\bauthor{\bsnm{Hricovini}, \binits{K.}},
\bauthor{\bsnm{Richter}, \binits{C.}},
\bauthor{\bsnm{Meyer}, \binits{J.C.}},
\bauthor{\bsnm{Kotakoski}, \binits{J.}},
\bauthor{\bsnm{Skákalová}, \binits{V.}}:
\batitle{Toward exotic layered materials: 2d cuprous iodide}.
\bjtitle{Advanced Materials}
\bvolume{34}(\bissue{9}),
\bfpage{2106922}
(\byear{2022})
{\href{https://arxiv.org/abs/https://onlinelibrary.wiley.com/doi/pdf/10.1002/adma.202106922}{{https://onlinelibrary.wiley.com/doi/pdf/10.1002/adma.202106922}}}.
\doiurl{10.1002/adma.202106922}
\end{barticle}
\endbibitem

\bibitem{Poikela_2014}
\begin{barticle}
\bauthor{\bsnm{Poikela}, \binits{T.}},
\bauthor{\bsnm{Plosila}, \binits{J.}},
\bauthor{\bsnm{Westerlund}, \binits{T.}},
\bauthor{\bsnm{Campbell}, \binits{M.}},
\bauthor{\bsnm{Gaspari}, \binits{M.D.}},
\bauthor{\bsnm{Llopart}, \binits{X.}},
\bauthor{\bsnm{Gromov}, \binits{V.}},
\bauthor{\bsnm{Kluit}, \binits{R.}},
\bauthor{\bparticle{van} \bsnm{Beuzekom}, \binits{M.}},
\bauthor{\bsnm{Zappon}, \binits{F.}},
\bauthor{\bsnm{Zivkovic}, \binits{V.}},
\bauthor{\bsnm{Brezina}, \binits{C.}},
\bauthor{\bsnm{Desch}, \binits{K.}},
\bauthor{\bsnm{Fu}, \binits{Y.}},
\bauthor{\bsnm{Kruth}, \binits{A.}}:
\batitle{Timepix3: a 65k channel hybrid pixel readout chip with simultaneous
  {ToA}/{ToT} and sparse readout}.
\bjtitle{Journal of Instrumentation}
\bvolume{9}(\bissue{05}),
\bfpage{05013}--\blpage{05013}
(\byear{2014}).
\doiurl{10.1088/1748-0221/9/05/c05013}
\end{barticle}
\endbibitem

\bibitem{gitlab}
\begin{botherref}
\oauthor{\bsnm{C.~Hofer}, \binits{T.J.P.} \bsuffix{C.~Gao}}:
PyPtychoSTEM.
GitLab
(2021)
\end{botherref}
\endbibitem

\bibitem{SUSI2019Efficient}
\begin{barticle}
\bauthor{\bsnm{Susi}, \binits{T.}},
\bauthor{\bsnm{Madsen}, \binits{J.}},
\bauthor{\bsnm{Ludacka}, \binits{U.}},
\bauthor{\bsnm{Mortensen}, \binits{J.J.}},
\bauthor{\bsnm{Pennycook}, \binits{T.J.}},
\bauthor{\bsnm{Lee}, \binits{Z.}},
\bauthor{\bsnm{Kotakoski}, \binits{J.}},
\bauthor{\bsnm{Kaiser}, \binits{U.}},
\bauthor{\bsnm{Meyer}, \binits{J.C.}}:
\batitle{Efficient first principles simulation of electron scattering factors
  for transmission electron microscopy}.
\bjtitle{Ultramicroscopy}
\bvolume{197},
\bfpage{16}--\blpage{22}
(\byear{2019}).
\doiurl{10.1016/j.ultramic.2018.11.002}
\end{barticle}
\endbibitem

\bibitem{Kirkland2010}
\begin{bbook}
\bauthor{\bsnm{Kirkland}, \binits{E.J.}}:
\bbtitle{{Advanced Computing in Electron Microscopy}},
\bedition{2}nd edn.
\bpublisher{Springer},
\blocation{New York}
(\byear{2010}).
\doiurl{10.1007/978-1-4419-6533-2}
\end{bbook}
\endbibitem

\bibitem{oleary2021increasing}
\begin{botherref}
\oauthor{\bsnm{O'Leary}, \binits{C.M.}},
\oauthor{\bsnm{Haas}, \binits{B.}},
\oauthor{\bsnm{Koch}, \binits{C.T.}},
\oauthor{\bsnm{Nellist}, \binits{P.D.}},
\oauthor{\bsnm{Jones}, \binits{L.}}:
Increasing spatial fidelity and snr of 4d-stem using multi-frame data fusion.
Microscopy and Microanalysis,
1--11
(2021).
\doiurl{10.1017/S1431927621012587}
\end{botherref}
\endbibitem

\bibitem{Langer2020}
\begin{barticle}
\bauthor{\bsnm{Langer}, \binits{R.}},
\bauthor{\bsnm{Błoński}, \binits{P.}},
\bauthor{\bsnm{Hofer}, \binits{C.}},
\bauthor{\bsnm{Lazar}, \binits{P.}},
\bauthor{\bsnm{Mustonen}, \binits{K.}},
\bauthor{\bsnm{Meyer}, \binits{J.C.}},
\bauthor{\bsnm{Susi}, \binits{T.}},
\bauthor{\bsnm{Otyepka}, \binits{M.}}:
\batitle{Tailoring electronic and magnetic properties of graphene by phosphorus
  doping}.
\bjtitle{ACS Applied Materials \& Interfaces}
\bvolume{12}(\bissue{30}),
\bfpage{34074}--\blpage{34085}
(\byear{2020})
{\href{https://arxiv.org/abs/https://doi.org/10.1021/acsami.0c07564}{{https://doi.org/10.1021/acsami.0c07564}}}.
\doiurl{10.1021/acsami.0c07564}.
\bcomment{PMID: 32618184}
\end{barticle}
\endbibitem

\bibitem{Mustonen2018}
\begin{barticle}
\bauthor{\bsnm{Mustonen}, \binits{K.}},
\bauthor{\bsnm{Hussain}, \binits{A.}},
\bauthor{\bsnm{Hofer}, \binits{C.}},
\bauthor{\bsnm{Monazam}, \binits{M.R.A.}},
\bauthor{\bsnm{Mirzayev}, \binits{R.}},
\bauthor{\bsnm{Elibol}, \binits{K.}},
\bauthor{\bsnm{Laiho}, \binits{P.}},
\bauthor{\bsnm{Mangler}, \binits{C.}},
\bauthor{\bsnm{Jiang}, \binits{H.}},
\bauthor{\bsnm{Susi}, \binits{T.}},
\bauthor{\bsnm{Kauppinen}, \binits{E.I.}},
\bauthor{\bsnm{Kotakoski}, \binits{J.}},
\bauthor{\bsnm{Meyer}, \binits{J.C.}}:
\batitle{Atomic-scale deformations at the interface of a mixed-dimensional van
  der waals heterostructure}.
\bjtitle{ACS Nano}
\bvolume{12}(\bissue{8}),
\bfpage{8512}--\blpage{8519}
(\byear{2018})
{\href{https://arxiv.org/abs/https://doi.org/10.1021/acsnano.8b04050}{{https://doi.org/10.1021/acsnano.8b04050}}}.
\doiurl{10.1021/acsnano.8b04050}.
\bcomment{PMID: 30016070}
\end{barticle}
\endbibitem

\end{thebibliography}






\clearpage

\section{Supplementary information}
\renewcommand{\figurename}{Supplementary Fig.}  

\setcounter{figure}{0}    

\subsection{Match between convolution method and quantitative simulated SSB image}
\begin{figure*}[!h]
	\center\includegraphics[width=0.85\textwidth]{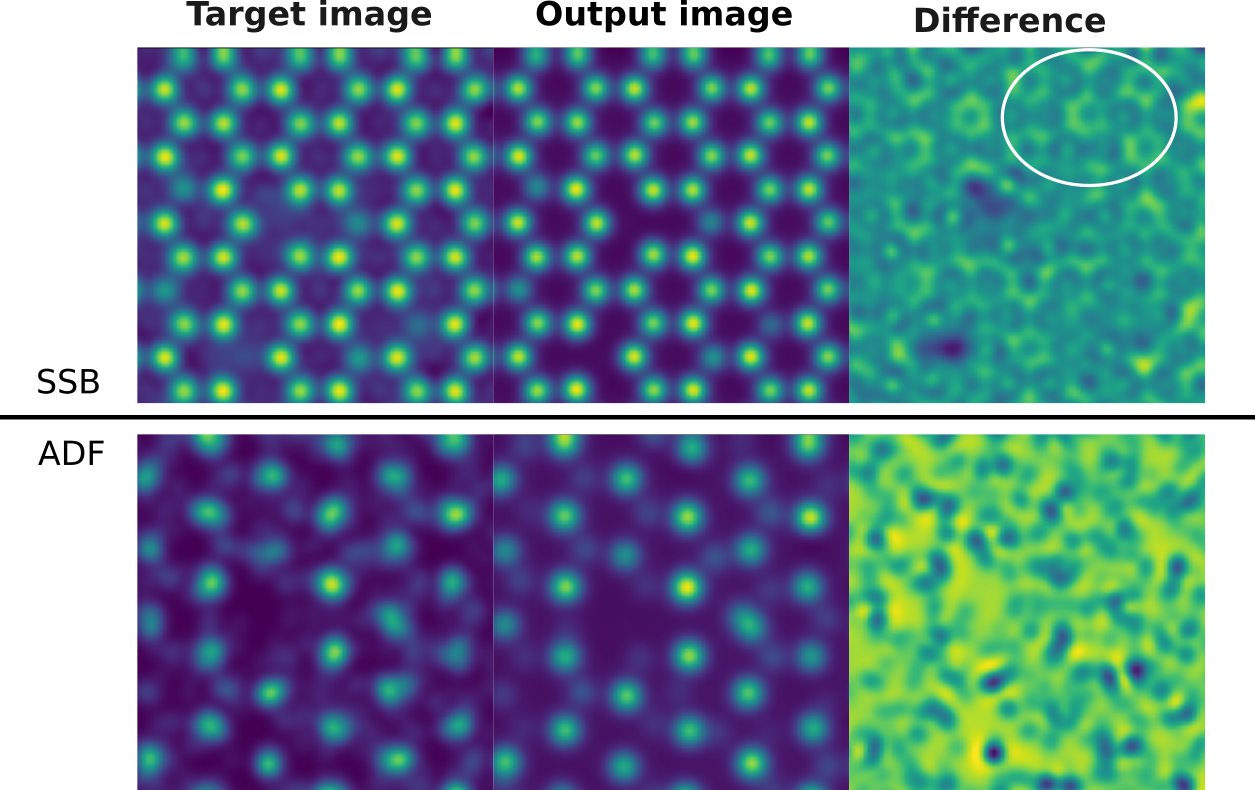}
	\caption{\label{Sfig:2D} Target image (quantitatively simulated by using the input model) of WS\textsubscript{2}, simulation of optimized model and pixel-by-pixel difference. The white circle indicates where the difference is the most pronounced.
	 }	
\end{figure*}

A convolution between the intensity profile of an electron probe and the potential of the structure can excellently reproduce ADF images of 2D materials as described in the main text. The contrast mechanism of phase images, as the SSB electron ptychography produces, are more complicated to simulate efficiently. However, the atomic positions and intensities are still well recovered by the optimization as Extended Data Fig.~\ref{Sfig:2D} shows. Here, the difference between the target image (which is the quantitatively simulated by the input model) and the optimized simulation still shows a pattern as a result of contrast reversals of the phase image.
More accurate simulation methods will be investigated in future work. 

\subsection{2D analysis}
\begin{figure*}
	\center\includegraphics[width=0.85\textwidth]{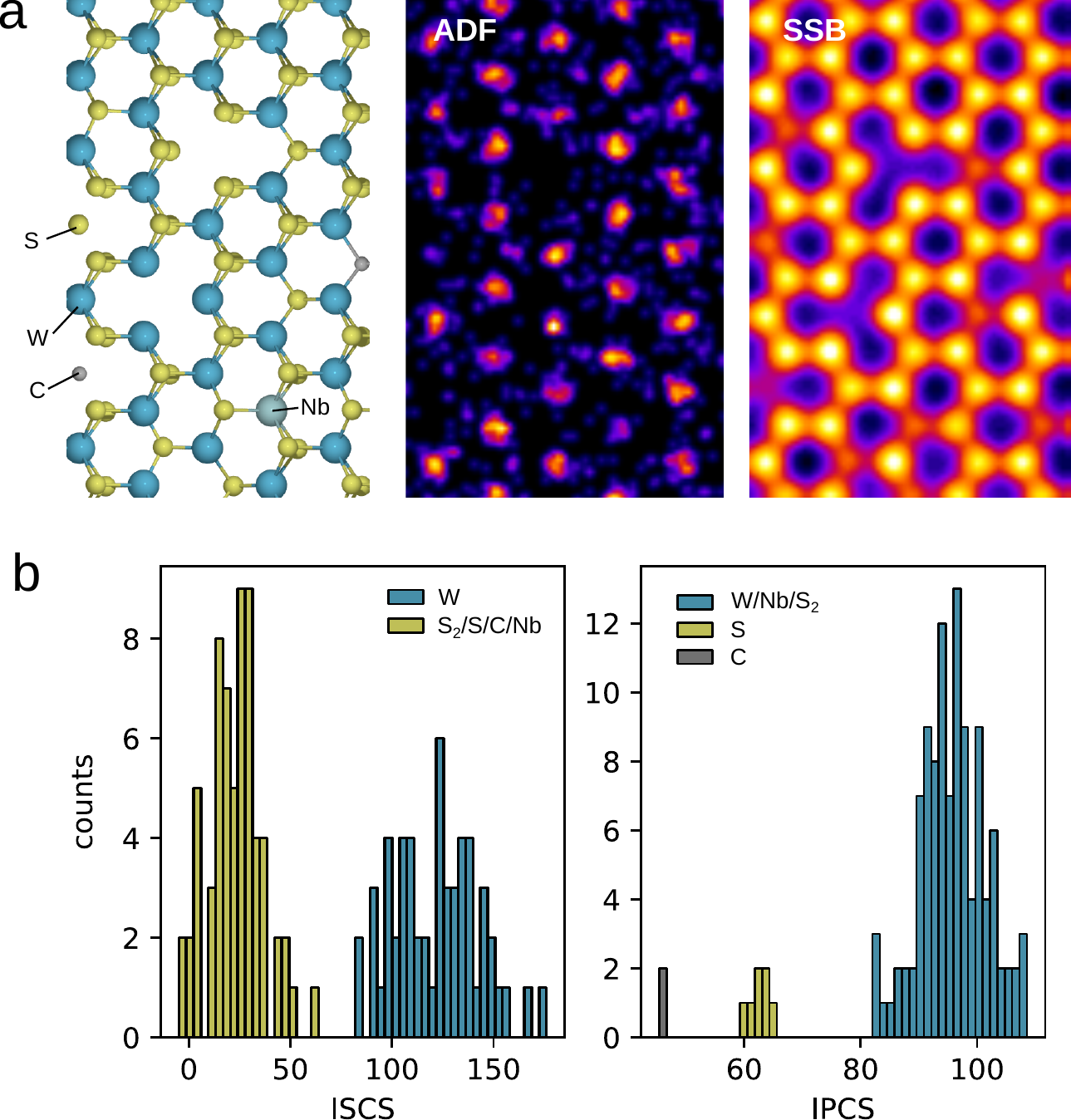}
	\caption{\label{fig:2D} \textbf{Analysis of the projected images.} (a) Artificial model of defective WS\textsubscript{2} (left), ADF simulation (center) and SSB simulation (right) at a dose of \SI{4000}{\electrons\per\angstrom\squared}. (b) Histogram of extracted atomic intensities using the ADF image (left) and the SSB image (right).
	}
\end{figure*}

Histograms of the extracted integrated scattering cross-sections (ISCS), obtained from the ADF image, and the integrated phase cross-sections (IPCS) obtained from the SSB data are shown in Extended Data Fig.~\ref{fig:2D}b. Two peaks are observed in the ADF ISCS histogram corresponding to the lighter S$_2$, S, C and Nb sites and the heavier W sites. The SSB IPCS shows three peaks distinguishing C, S and the other sites. Since the phase image also contains a negative contribution, the squared phase~\cite{oleary2021increasing} could be considered instead of the IPCS for the analysis. However, we abstain from taking the square since our analysis already reliably reproduces the chemical elements from the raw SSB image. 

The ptychographic phase and ADF signals compensate each other's weaknesses for elemental characterization (cf.\ Extended Data Fig.~\ref{fig:2D}a). 
While the ADF easily distinguishes the W and the S\textsubscript{2} sites, the signal is not sufficient to directly extract single S vacancies and other light elements from the S\textsubscript{2} site at this dose. The SSB image however is capable of distinguishing the lighter elements such as S and C, despite the neighboring heavy W sites. 
For this analysis, we consider the ISCS and IPSC as values which are capable of distinguishing the \textit{a priori} known elements.
The identification of dopants sometimes requires assumptions about their expected position within the unit cell. For example, Nb can be identified because its scattering intensity is significantly weaker than the W intensity and it is located at the metal site.
Note that the element has to be known and Nb is difficult to be distinguished from S\textsubscript{2} either by ADF or SSB, but still can be unambiguously assigned as it is positioned in the W-lattice site.
Under very high doses, ADF imaging would in principle be able to distinguish these elements, but we want to avoid high electron exposures here as we want to spend the dose budget for different beam incident angle acquisitions and generally avoid damaging the structure. 

\subsection{Details of the algorithm}
\begin{figure*}
	\center\includegraphics[width=0.8\textwidth]{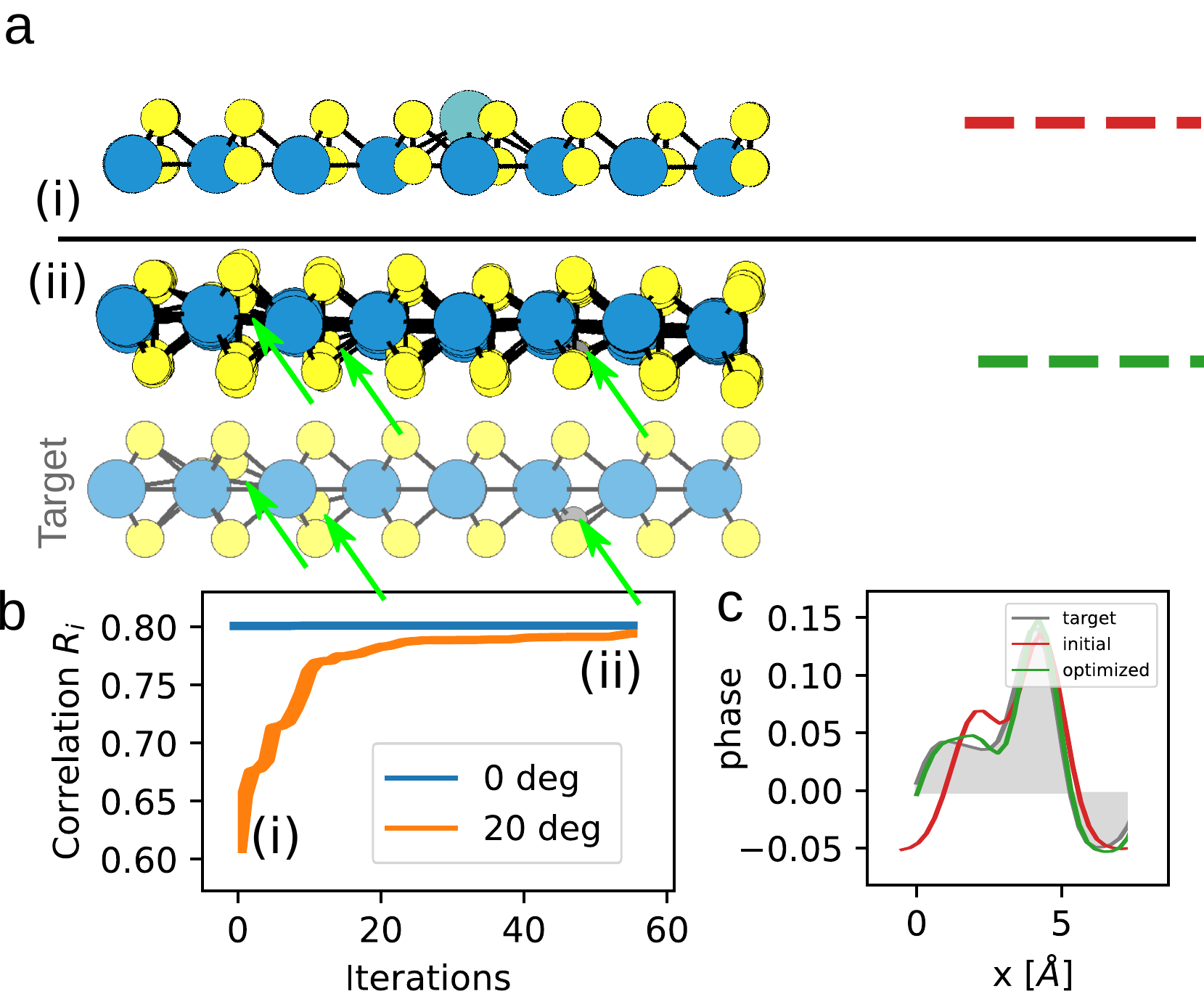}
	\caption{\textbf{\label{fig:conv} Progress of the algorithm using two tilted projection as an example} (a) Initial (i) and optimized (ii) model with the tilted simulations. (b) Plot of correlation as a function of iteration of 0 and 20 degrees. (c) Line profile of the simulations in a and the target image.
	}	
\end{figure*}

The reconstruction is illustrated in Extended Data Fig.~\ref{fig:conv}. Panel a shows the initialized model with incorrect z coordinates with the corresponding 20$^\circ$  tilt SSB simulation of WS\textsubscript{2}. During optimization, the correlation of the simulation corresponding to the updated model with the target image increases (panel b). A line profile of the simulation at the initial (red) and optimized (green) model reveals an excellent match of the latter one with the target image (see panel c). The reconstructed model in panel a also looks visually very similar to the target model, including all of the defects (indicated by the green arrows).

\subsection{Estimation of accuracy}
\begin{figure*}
	\center\includegraphics[width=1\textwidth]{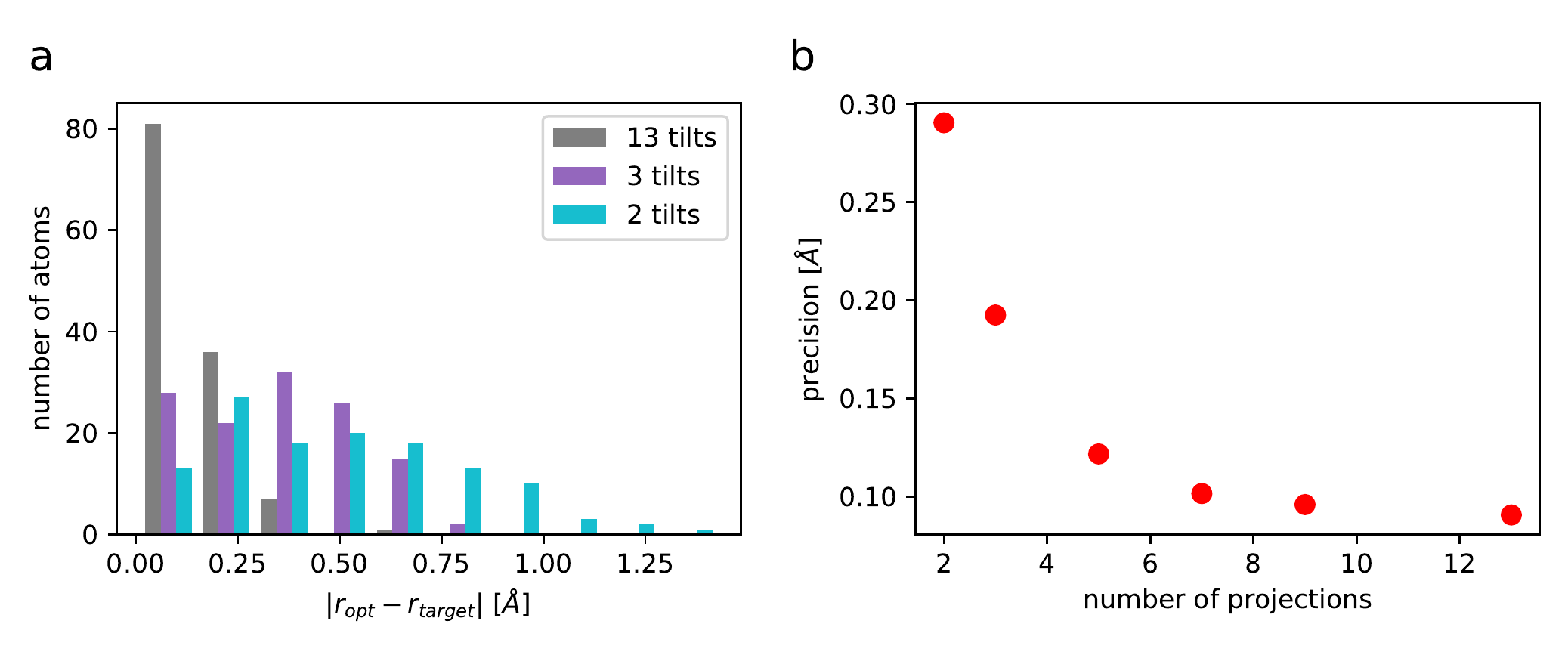}
	\caption{\label{fig:acc} Histogram of absolute distances between the corresponding atoms of the input model (WS\textsubscript{2}) and the optimized model using two, three and 13 tilts, respectively. Each (tilted) projection is based on an electron dose of \SI{4000}{\electrons\per\angstrom\squared}. The 13 tilts range from 5--30 degrees in both tilt axis.
	}	
\end{figure*}

The predefined target model allows us to quantify the accuracy and precision of the method by calculating the difference between the target model and the optimized model. The absolute difference between the corresponding atoms in the two models is summarized in Fig.\ \ref{fig:acc}a showing a histogram of the differences using different numbers of tilt angles and the estimated precision via the standard deviation of the corresponding histograms. The total precision (including C,Nb, S and W) ranges from 9--29 pm depending on the number of projections. This demonstrates that picometer precision is achievable with very few tilts with total electron doses below those typical for imaging 2D materials. Given these numbers, the method is capable of obtaining the precision with only a 30th dose than other tomographic approaches~\cite{Tian2020}.

\subsection{Precision analysis of CuI}
\begin{figure*}[!h]
	\center\includegraphics[width=0.95\textwidth]{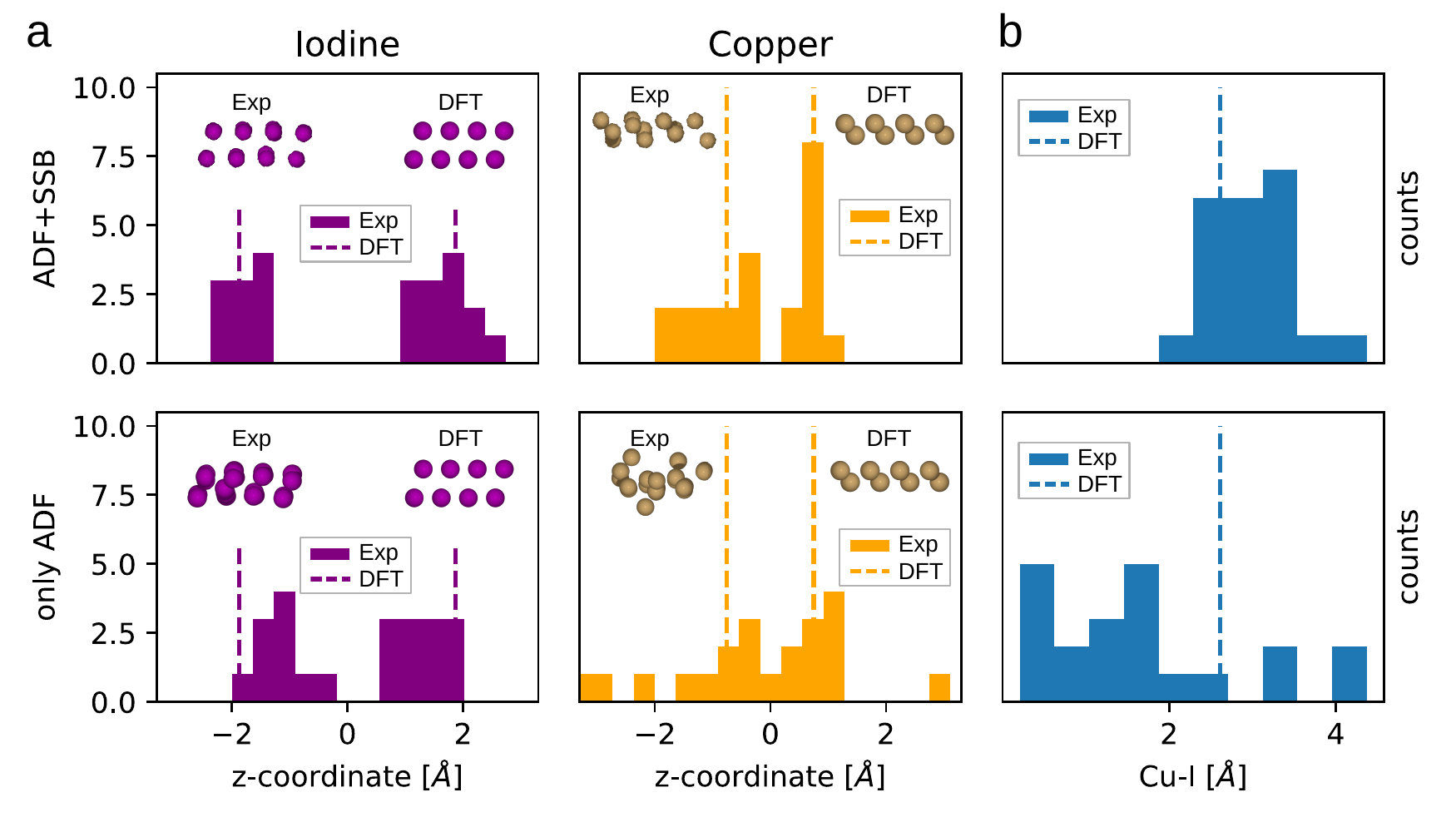}
	\caption{\label{fig:hists}\textbf{Analysing the precision of the reconstruction of CuI.} 
	a) Histogram of optimized z positions of iodine and copper (left and right, respectively) using both ADF and SSB and only ADF (top and bottom, respectively). The corresponding atomic models are schematically shown as insets to the plots. b) Histogram of the distances between iodine and Cu in the same lattice site using both ADF and SSB images for each projection (left) and only the ADF images (right) respectively.
	 }	
\end{figure*}

We now further analyse the precision of the experimental reconstructions and quantitatively demonstrate the utility of electron ptychography for accurate reconstructions at low doses. Because the models are purely based on experimental data, the real structure is not known. However, the distribution of atomic heights reveals two separate peaks for the iodine and copper atoms representing the average height of both sublattices as seen in Extended Data Fig.\ \ref{fig:hists}. Their mean values excellently agree with DFT calculations (dashed lines in Extended Data Fig.\ \ref{fig:hists}) when using both ADF and SSB together for the reconstruction. 
The standard deviations of the distribution of absolute height differences are calculated to be \SI{12.8}{\pm} and \SI{14}{\pm} for iodine and copper, respectively. The in-plane coordinates are neglected by this method, justified by the much higher precision of determining the projected (x,y) positions. The precision of the experimental reconstruction is slightly less than would seem indicated by the WS$_2$ simulations. This discrepancy can be explained by the presence of residual non-linear drift, scan distortions, the finite accuracy of the tilting holder, contamination and out-of-plane atomic vibrations.

When excluding the SSB images, the reconstructed iodine positions not only have a broader distribution with a standard deviation of \SI{37}{\pm}, but the distance between both planes is also smaller than expected from DFT calculations. We assume that this error comes from the convolution with the incorrectly assigned Cu intensities. The precision of the Cu positions is, as expected, much poorer with a standard deviation of \SI{62}{\pm}.

We also analysed the distances between Cu and I of the same lattice sites of both optimized models (using only ADF and ADF and SSB, respectively). Also here, the combination of SSB and ADF images produces the most reliable reconstruction with a precision of \SI{27}{\pm} which agrees with our previous analysis. 
Excluding the SSB in the optimization drastically degrades the precision to \SI{73}{\pm}.

Few tilt tomography using only ADF images has been used with graphene to reveal out-of-plane distortions of defects and  distortions in graphene~\cite{Hofer2018,Hofer2019,Langer2020,Mustonen2018}. However, CuI consists of four atomic layers rather than a single layer. One could imagine that a similar precision could be obtained by increasing the number of ADF images taken at different tilt angles with this more complex structure. However, especially as a very high signal-to-noise ratio is required in order to reveal the contrast of Cu, this has the major drawback of a significantly higher dose requirement. As we have shown, electron ptychography is an excellent method to overcome those issues: At low doses, the quality of the reconstructed images are significantly improved and it reveals the locations of more elements including the projected positions of the light elements. 

The analysis can be further extended to the refinement of the 3D atomic positions by only using the phase image, since the SSB image has usually a significantly better signal-to-noise ratio than ADF images. Therefore, one might expect that including the ADF might lead to a less accurate reconstruction than only using the SSB.   
Our reconstructions, however, show that the exclusion of the ADF leads to a poorer structural refinement. We assume that the high intensity of I in the ADF signal more than compensates for the noisy Cu signal in the ADF images as noise does not have a high contribution to the change of the correlation function. 

\subsection{Long-range curvature}
\begin{figure*}[!h]
	\center\includegraphics[width=0.8\textwidth]{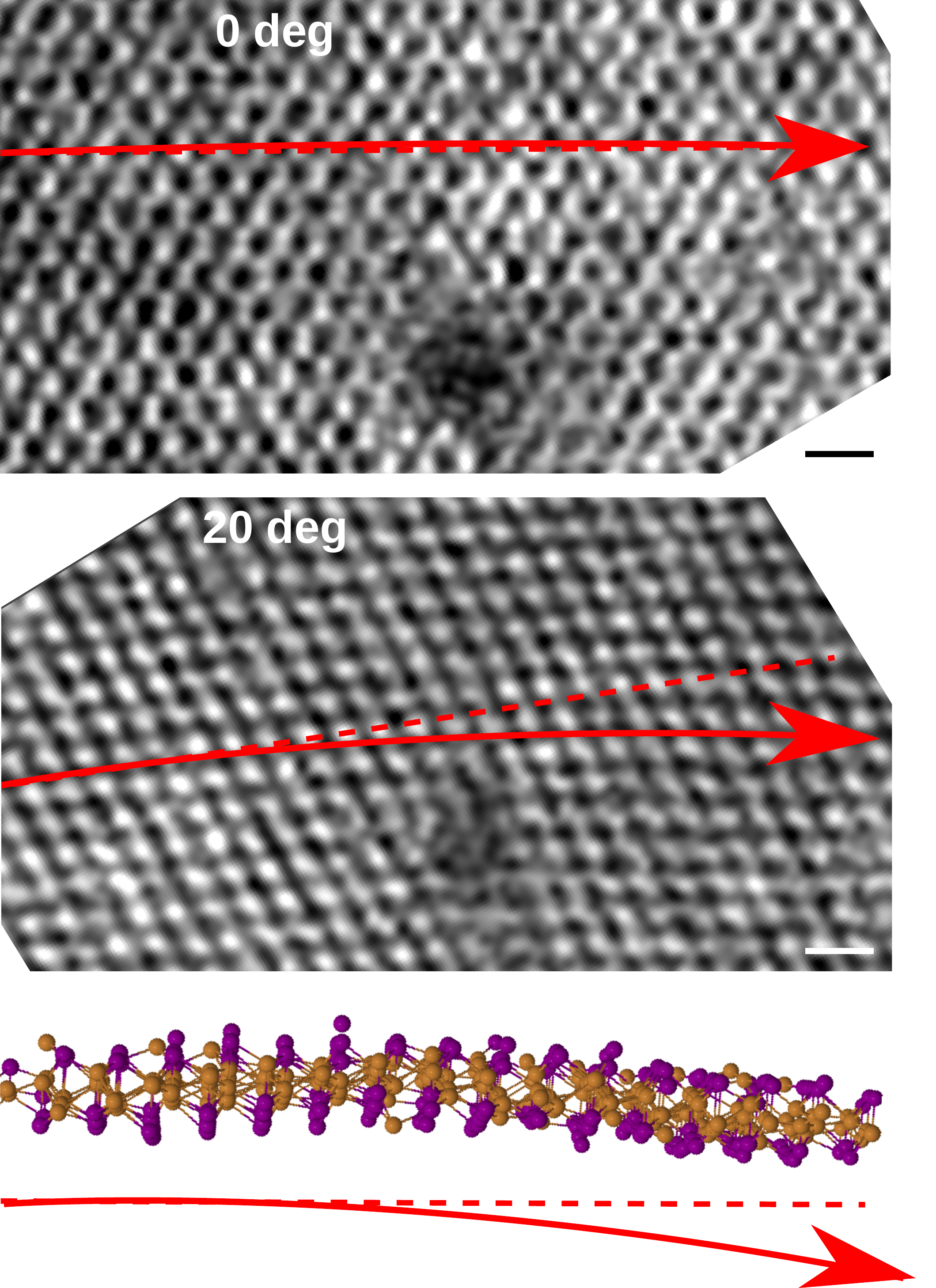}
	\caption{\label{Sfig:curve} \textbf{Long-range 3D reconstruction of 2D CuI showing a significant lattice curvature.} The 20 degrees images (here we only display the ADF) show a lattice distortion compared to the 0 tilt images. This distortion is directly related to a 3D deformation as shown in the model below, as a result of a symmetry break in the tilted projection. }	
\end{figure*}

Extended Data Fig.~\ref{Sfig:curve} shows that 2D CuI is not necessarily flat. Although only two tilts were acquired from this specific region, the reconstruction shows a significant undulation of the structure. We can exclude that this is an artifact from the smaller precision as the lattice distortion is already visible in the tilted projection as indicated by the red arrows in Extended Data Fig.~\ref{Sfig:curve}. This shows that even low-frequency features in 2D materials such as flake curvatures can be easily reproduced by only two tilt tomography.

\subsection{Complete few-tilt series}
\begin{figure*}[!h]
	\center\includegraphics[width=1\textwidth]{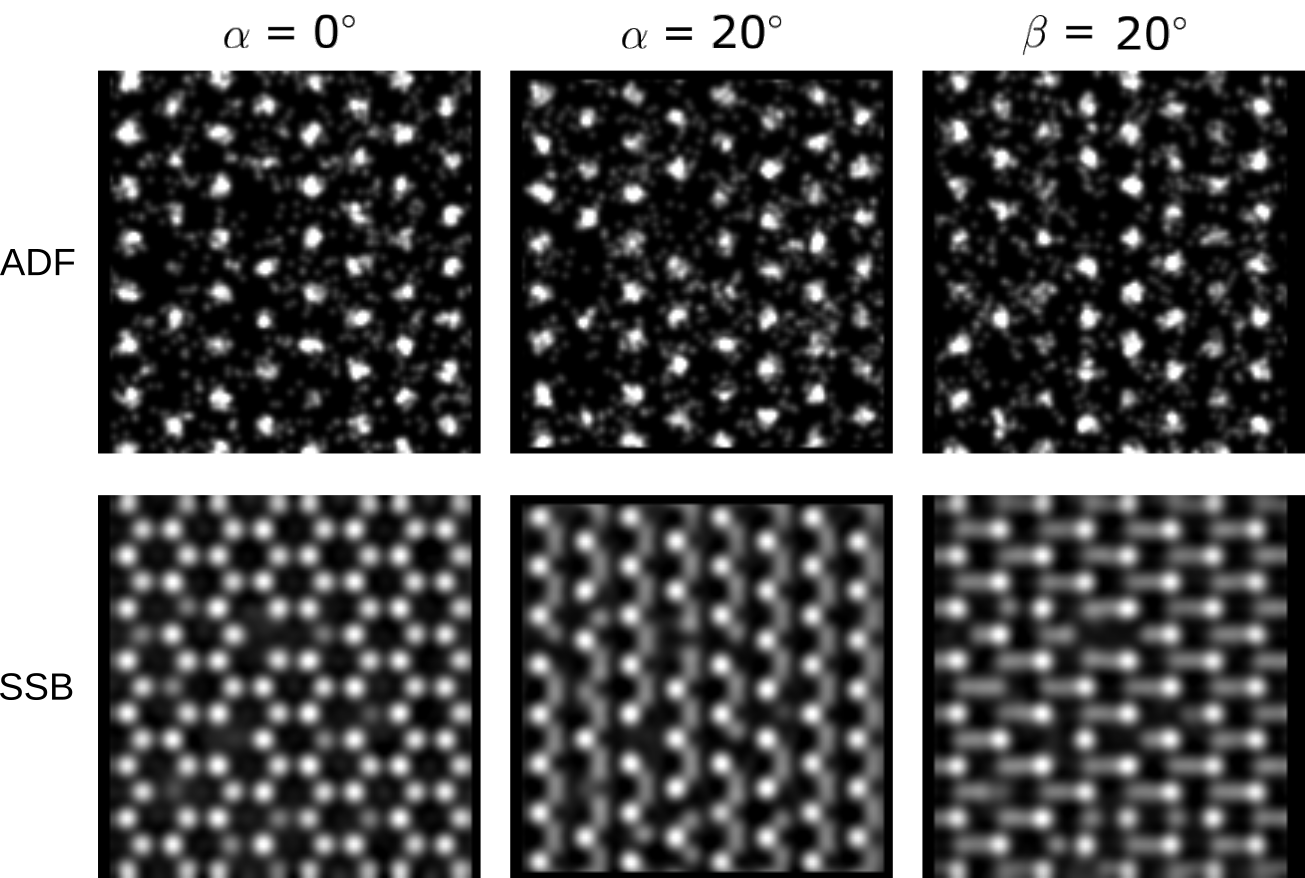}
	\caption{\label{Sfig:raw2} \textbf{Three-tilt tomographic series of WS\textsubscript{2}.} Simulated ADF (top) and SSB phase images (bottom) for different tilt angles. The simulation parameters are described in the methods part of the main text of the  manuscript. 
	 }	
\end{figure*}

Extended Data Fig.~\ref{Sfig:raw2} shows the simulated ADF and SSB images used for the 3D reconstruction of WS\textsubscript{2}. The dose for each image is \SI{4000}{\electrons\per\angstrom\squared}. 

\begin{figure*}[!h]
	\center\includegraphics[width=1\textwidth]{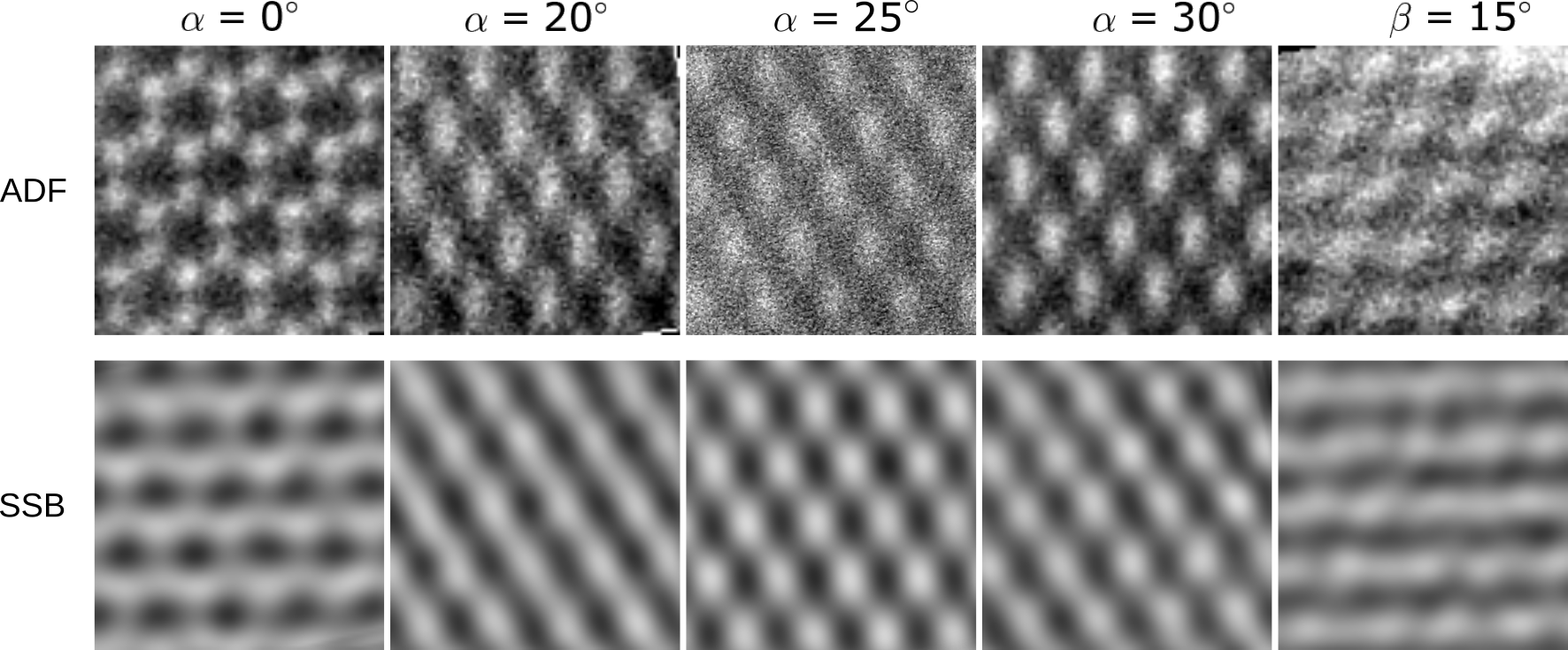}
	\caption{\label{Sfig:raw} 
	 \textbf{Five-tilt tomographic series of CuI.} ADF (top) and SSB phase images (bottom) for different tilt angles which are used for the reconstruction of the model. The acquisition parameters are described in the methods part of the main text of the  manuscript. }	
\end{figure*}

 For the experimental data set, we reduced the computational expense by cropping the full $1024\times1024$ image into 16 $256\times256$ images and then dropped out the images which were visibly too contaminated. We then aligned and averaged the other images, which is possible as we assume periodicity of the crystal over the whole field-of-view. Apart from a further improvement of the signal-to-noise ratio of the final image, the number of atoms in the system is also reduced. Extended Data Fig.~\ref{Sfig:raw} shows the ADF and SSB images used for the 3D reconstruction of CuI. The dose of each image is estimated to be approx. \SI{20000}{\electrons\per\angstrom\squared}.  This is indeed sufficient to obtain a decent signal-to-noise image of both methods.

\end{document}